\documentclass[onecolumn,
 reprint,
 amsmath,amssymb,
 aps,
groupedaddress,superscriptaddress]{revtex4-2}
\usepackage[normalem]{ulem}
\usepackage{graphicx}
\usepackage{dcolumn}

\usepackage{siunitx}
\usepackage{color}
\usepackage{amsmath}

\begin{document}
\preprint{APS/123-QED}

\title{Enantioselective optical forces in gain-functionalized single core-shell chiral nanoparticles}

\author{R. Ali}
\email[]{rali.physicist@gmail.com}
\affiliation{%
Applied Physics Department, Gleb Wataghin Physics Institute, University of Campinas, Campinas 13083-859, SP,
Brazil
}
\author{F. A. Pinheiro}
\affiliation{Instituto de F\'isica, Universidade Federal do Rio de Janeiro, Caixa Postal 68528, Rio de Janeiro, Rio de Janeiro 21941-972, Brazil}

\author{R. S. Dutra}

\affiliation{%
{LISComp-IFRJ },  
{Instituto Federal de Educa\c c\~ao, Ci\^encia e Tecnologia, Rua Sebasti\~ao de Lacerda, Paracambi, RJ, 26600-000, Brasil }
}
\author{T. P. Mayer Alegre}
\affiliation{%
Applied Physics Department, Gleb Wataghin Physics Institute, University of Campinas, Campinas 13083-859, SP,
Brazil
}
\author{G. S. Wiederhecker}
\email[]{gsw@unicamp.br}
\affiliation{%
Applied Physics Department, Gleb Wataghin Physics Institute, University of Campinas, Campinas 13083-859, SP,
Brazil
}

\begin{abstract}
We propose a gain-assisted enantioselective scheme in dye-doped chiral particles, demonstrating optical pulling and pushing forces that can be tuned using externally controllable parameters. By controlling the concentration of dye molecules and pumping rate, we achieve all-optical chiral resolution of racemic mixtures and enantioselection of small dipolar chiral particles without relying on interference. This scheme is applicable to both lossless and lossy plasmonic chiral particles, making it a promising approach for chiral sensing, drug discovery, and molecular separation.
\end{abstract}

\maketitle

\maketitle
\section{Introduction}
An object is said to be chiral if it has distinguishable mirror images, these two opposite mirrored forms are called enantiomers \cite{Wagniere2007,Hentschel2017,fan2010,kuzyk2012,zhang2005}.
The separation of chiral enantiomers is an important scientific and technological task with many multidisciplinary applications \cite{Brooks2011,Urban2019}.
In the last decades, considerable effort and progress have been made on the development of highly efficient optical enantioselective methods,
such as enantioselective optical trapping \cite{dionne2016,Zhao2017,Tkachenko2014Optofluidic}, oppositely directed lateral forces \cite{wang2014,chen2016,zhang2017}, enantioselective pulling forces \cite{Ali2021,Zheng2021Light}, and azimuthal/longitudinal optical torques \cite{Ali2020nanoscale,Manman2021}. 
Despite the rich literature on optical enantioselective schemes, they typically involve the trapping of a lossless particle using a structured optical beam or tightly focused beam. The enantioselection of the lossy  chiral sphere and chiral dipoles regardless of absorption always remain challenging and restricted due to the weak restoring force.   
This restriction occurs because optical trapping and pulling forces rely on the relative strength between optical restoring (coming from the intensity gradient or interfaces between the incident and scattered field) and scattering forces (coming from the optical scattering). When considering more general lossy particles, such force competition  strongly depends both on  absorption level and  particle size. As a result, regardless of the particle size, enantioselection becomes difficult  due to strong pushing forces, which increase with absorption. 
In addition, in the case of a dipolar chiral particle with a negligibly small magnetic dipole, the interface between the dipoles is small, therefore,  the realization of the pulling effect is impossible due lack of the restoring force.

To circumvent these limitations, we propose an alternative strategy that simultaneously enables chiral separation and long-range optical manipulation of a single chiral sphere. We consider an optical gain-functionalized chiral sphere, which enables one to exert tunable optical pulling and pushing forces using a single circularly polarized (CP) plane wave. The main advantage of using gain is to eliminate the dependence on interference to achieve the restoring forces, in contrast to existing enantioselective methods based on optical forces \cite{Ali2021,wang2014,Hayat2015,chen2016,Chen2011optical}. Here, restoring force is achieved through the emission of photons due to gain mechanisms, and the number of emitted photons increases with the active size of the sphere. As the optical gain can be experimentally controlled with great success, it allows us to exert the necessary tunable optical pulling and pushing forces on-demand.
 This work is intended to carry out a detailed study of the enantioselection of  chiral Mie spheres and chiral dipoles by employing experimentally controllable parameters such as the laser pump rate of the dye molecules. 
 
 To accomplish this study, first, we consider a chiral core-shell particle to establish the underlying physics and connection between the gain mechanism and optical forces. Further, we explore 
 the enantioselection of the chiral particles using experimentally achievable parameters. After establishing the enantioselection with the gain media we check the  viability and robustness of this scheme by discussing the optical forces acting on the  homogeneous chiral sphere and chiral dipoles. In addition, we also show that our scheme can perform enantioselection of the chiral molecules  with arbitrarily  small chiral parameters and small sizes.

\section{Results and discussion}
\subsection*{Optical force on chiral core-shell}
To begin with, we consider  chiral core-shell spheres to calculate the chiral and optical gain-dependent optical forces. This model is general enough to describe a wide range of chiral nanoparticles of scientific and technological interest, such as gain-doped chiral nanocrystals~\cite{fan2012chiral}, and core-shell particles with a dielectric core and a plasmonic shell
\cite{Ali2020nanoscale,Lu2018,Rao2015,Lan2016,Cipparrone2011}.
To the sake of generality in the Appendix we consider the case of a single homogeneous chiral sphere doped with gain, a model that can be applied to gain-doped chiral nanocrystals~\cite{fan2012chiral}. To this end let us consider a CP plane wave ${\bf E} = E_0 (\hat{x}+i\sigma \hat{y})e^{ ik z}$ ($\sigma=+1 (-1)$ is the spin index of left- (right-) handed CP plane wave) impinging on particles made of dielectric core (radius $b$, relative permittivity $\epsilon_{\text{d}}$) and a chiral shell 
 of thickness $t$, refractive index $n_\sigma={\epsilon_s}+\sigma \kappa$, where $\epsilon_s$ is relative permittivity and  $\kappa$ is the chiral index, immersed in water with relative permittivity ${ \epsilon_m = 1.77}$.  
The optical force acting on the sphere due to a plane wave is written in terms of scattering efficiency $Q_\text{s}$ and absorption efficiency $Q_\text{a}$ as~\cite{Bohren}
\begin{equation}
F = \frac{\epsilon_0 \epsilon_m E_0^2}{k^2_m}(Q_\text{a}+ Q_\text{s} (1 - \langle\cos\theta\rangle))= F_\text{a} + F_\text{s} \label{force}
\end{equation}
where $E_0$ is electric field amplitudes, ${ \epsilon_0}$ is the vacuum permittivity, $k_m$ is wave vector inside the surrounding medium and $<\cos\theta >\equiv g$ is the scattering asymmetry parameter~\cite{Bohren} ($\theta$ is the scattering angle). The total force can be divided into two force contributions: scattering force $F_\text{s}$ and $F_\text{a}$ absorption force~\cite{Bohren}. 
 It is clear from Eq. \ref{force} that for a passive chiral sphere, where $\{Q_{a}, Q_s\}\geq0 $, $-1\leq g \leq 1$, $\{F_{a}, F_s\}\geq0 $, the optical force $F$ is always positive, regardless of the chiral handedness, in which case chiral resolution impossible. 

 \begin{figure}
\centering
\includegraphics[scale=1.05]{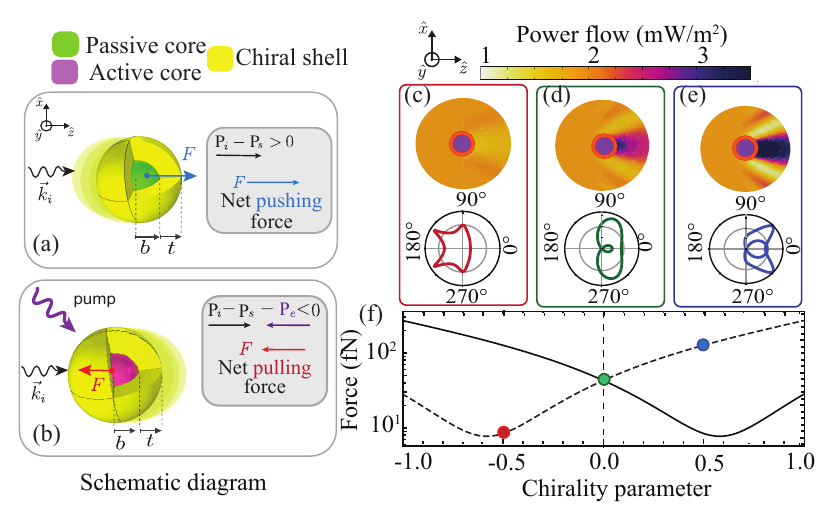}
\caption{Schematic of optical forces on the sphere: (a) a CP plane wave scattering by core/shell, passive core and chiral shell, particle, where $\textbf{P}_i$ and $\textbf{P}_s$ are the momenta of the incident and scattered photons. The resultant net optical force on the sphere is pushing $F \geq 0$. (b) Doped core, dye molecules, and pump laser enable  a gain mechanism, as a result, photons are emitted with momentum $\textbf{P}_e$ and the net optical force on the active chiral core-shell can be negative. 
Numerically calculated power flow of the relative scattered fields and polar scattering pattern by the passive chiral core/shell for chirality parameter (c) $\kappa=-0.5$, (d) $\kappa=0$ and (e) $\kappa=0.5$. (f) The optical force acting on the passive sphere as a function of the chirality parameter for left CP (dashed) and right CP (solid).} 
 \label{F1} \end{figure}

However, in the presence of the gain media, the momentum of stimulated emission photons may be strong and one may expect $F_\text{s}>0$ and $F_\text{a}<0$, hence a fine-tuning of the gain level allows to control the direction of $F$ on each chiral face, as indicated in Fig.~\ref{F1}(a,b). 
For instance,  a passive particle, Fig.~\ref{F1}a, the momentum of the scattered photon $\textbf{P}_s$ can never be larger than the momentum of incident photon $\textbf{P}_i$ (i.e. $\textbf{P}_i-\textbf{P}_s\geq0$). Therefore, the momentum transfer to the particle should lead to a net positive (pushing) force. In contrast, in an active particle with sufficient gain, Fig.~\ref{F1}b, $F_a$ may have its sign reversed as it may be dominated by the recoil from stimulated emission photons with momentum $\textbf{P}_e$ \cite{Mizrahi2010,Kudo2012}, inducing a negative (pulling) force $F_a$, as illustrated in Fig.~\ref{F1}(b). Since the shell is chiral, in addition to the gain-enable pulling force $F_\text{a}<0$, the particle refractive index depends on the incident field helicity, which leads to a distinct scattering efficiency $Q_s$ for each enantiomer. Therefore, the gain-tunable pulling force -- associated with the chiral-sensitive pushing scattering force -- is the key mechanism underlying our tunable enantioselection scheme. It is worth mentioning that the force contribution of the pump laser can be ignored by considering its normal orientation  to the incident laser axis. 

 We quantitatively demonstrate the effectiveness of our enatioselection scheme by considering a passive core ($\epsilon_=2.5$) radius $b=\SI{ 200}{\nano\meter}$, and chiral shell ($n_\sigma=1.7+\sigma\kappa$) of thickness $t=\SI{ 80}{\nano\meter}$ illuminated by a left-CP plane wave. Using COMSOL Multiphysics $^\text{\textregistered}$ we numerically calculate the power flow of the scattered fields for different chirality parameters $\kappa=-0.5$, $\kappa=0$ and $\kappa=0.5$ as shown in Fig. \ref{F1}(c), \ref{F1}(d) and \ref{F1}(e), respectively. The bottom row of Figs. \ref{F1}(c)-(e) shows the polar representation of the scattered Far field. It is clear that the scattered momentum for the left-handed chiral sphere in Fig.\ref{F1}(e) is larger than that of the right-handed sphere in Fig.\ref{F1}(c), in agreement with the qualitative arguments above. When particle and beam handedness are opposite (lower refractive index), scattering is minimum and hence the pushing optical force is minimized, as shown in Fig \ref{F1}(f), where the optical force as a function of the chirality is shown for right (dotted-line) and left (solid-line) -handed CP incident light. 

\begin{figure}
\centering
\includegraphics[scale=1]{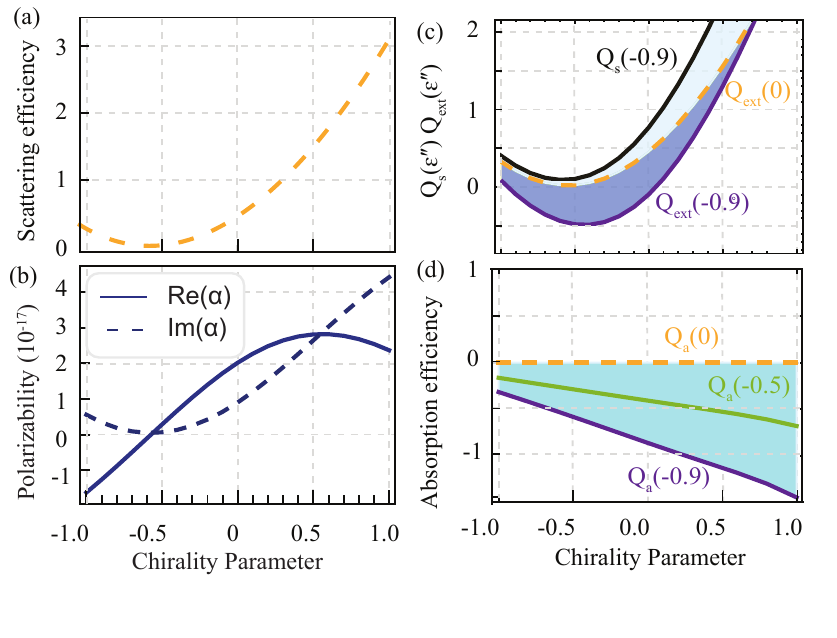}
\caption{ (a) Scattering efficiency $Q_{s}$ and (b) total chiral polarizability of the passive chiral sphere as function of chirality parameter. (c) Scattering efficiency $Q_{s}(\epsilon")$, extinction efficiency $Q_{ext}(\epsilon")$ and (d) absorption efficiency versus chirality parameter $Q_{a}(\epsilon")$ for different imaginary permittivity $(\epsilon")$ (controlled by the gain parameter), remaining parameters are the same as in Fig.\ref{F1}} 
 \label{F2}
\end{figure}
To understand the precise role of the chiral refractive index in the scattering force, we show the scattering efficiency in Fig. \ref{F2}(a) (left CP incident beam). Indeed, $Q_s$ is very weak at $\kappa=-0.5$ and a stronger $Q_s$ is found for $\kappa=0.5$. This behavior is also consistent with chiral sphere polarizability $\alpha$, as shown in Figure \ref{F2}(b). The magnitude of both the real and imaginary parts are minimized at $\kappa=-0.56$, strongly suppressing the scattered radiation and, thus, revealing the physics underlying the suppression of the scattering force.
 In contrast, the sphere with opposite chirality, $\kappa=+0.56$, has a large polarizability and hence a large scattering efficiency and pushing force.

Insofar the chiral response of our particles is solely influenced by the passive optical properties. We now discuss the impact of the optical gain on optical forces. In Fig.\ref{F2}(c) we calculate $Q_\text{ext}(\epsilon", \kappa)$ \cite{Bohren8} and $Q_\text{s}(\epsilon", \kappa)$ of the particles as a function of the chirality parameter with fixed real relative permittivity $\epsilon_{\text{d}}$ and for different imaginary permittivity $\epsilon"$ of the core -- which can be controlled by the pump-induced gain level; the other geometric parameters are the same as in Fig. \ref{F2}(a). While $Q_\text{ext}=Q_\text{s}$ for a passive core, this identity does not hold for an active core particle. In this case, the extinction efficiency becomes negative, for larger gain, for the right-handed chiral sphere ($Q_\text{ext}(-0.9,\kappa<0)$ in Fig.\ref{F2}(c)). This difference appears in the absorption efficiency $Q_{a}(-0.9, \kappa)=Q_{ext}(-0.9,\kappa) - Q_{s}(-0.9, \kappa)$. Note that for a lossy sphere extinction efficiency is always less than scattering efficiency and $Q_a$ remains positive and optical forces on the sphere are pushing forces. However, for the active sphere, the pump rate of the gain medium allows us to achieve negative absorption and the corresponding optical force becomes a pulling force.  Figure \ref{F2}(d) shows that for larger gain level $Q_{a}(\epsilon", \kappa)<0$ for a broad range of values of $\kappa$, so that by Eq. \ref{force} optical pulling force ($F<0$) can be achieved. 

To ground our study in specific context, we consider a realistic gain model where the dielectric core is doped with a judicious amount of dye molecules \cite{Pezzi2019,Polimeno2020,Campione2011}.
For our purpose, we consider the gain properties of solvatochromic LDS 798 dye molecules \cite{Pezzi2019,Doan2017,Ali2022acs} in a dielectric $\epsilon_{\text{d}}$ host, which is a well established combination to achieve active particles \cite{Shi2022,Ali2021jopt}.
The dye molecules are modeled as four-level atomic systems with occupation number density $N_i$, and $\sum_0^4 N_i=N_\text{dye}$ represents the total number of dye molecules per cubic meter. The effective permittivity of the dye-doped dielectric core is~\cite{Campione2011,Polimeno2020,Pezzi2019} 
\begin{equation}
\epsilon_{g} = \epsilon_0\epsilon_{\text{d}} + \frac{\sigma_b\, N_\text{dye}}{\omega^2+i\Delta\omega_a\omega-\omega_a^2}\frac{(\tau_{21}- \tau_{10})\Gamma_\text{pump}}{1+(\tau_{32}- \tau_{21}+\tau_{10})\Gamma_\text{pump}}, \label{neff}
\end{equation}
where $\sigma_b= \frac{6 \pi \epsilon_0c^3\eta}{\tau_{21}\omega_a^2 \sqrt{\epsilon_h}}$, $\omega = 2\pi c/\lambda $, $c$ is the speed of light, $\Delta\omega_a$ is the bandwidth of the dye transition between the two levels resonant with the incident beam, $\omega_a = 2\pi c/\lambda_a $ is the frequency of the emitted photon, $\tau_{i+1,i}$ is the relaxation time between the $(i+1)^\text{th}$ and $i^\text{th}$ level $\Gamma_\text{pump}$ is the pump rate of the dye molecules \cite{Pezzi2019,Doan2017}. 
Equation \ref{neff} indicates that the optical gain can be externally controlled by appropriately tuning the emission photons by means of $N_{dye}$ and the pump rate of the dye to activate the gain mechanism. 
We choose the following parameters to describe: $\Delta\omega_a =175$ THz, $\lambda_a = 777 { \,{\rm nm}}$, $ \eta = 0.48$, ${\tau_{10}= \tau_{32}=100 {\rm\, fs}}$ and ${\tau_{21} = 50 \,{\rm ps}}$ \cite{Pezzi2019,Doan2017}.

\begin{figure}
\centering
\includegraphics[scale=.4]{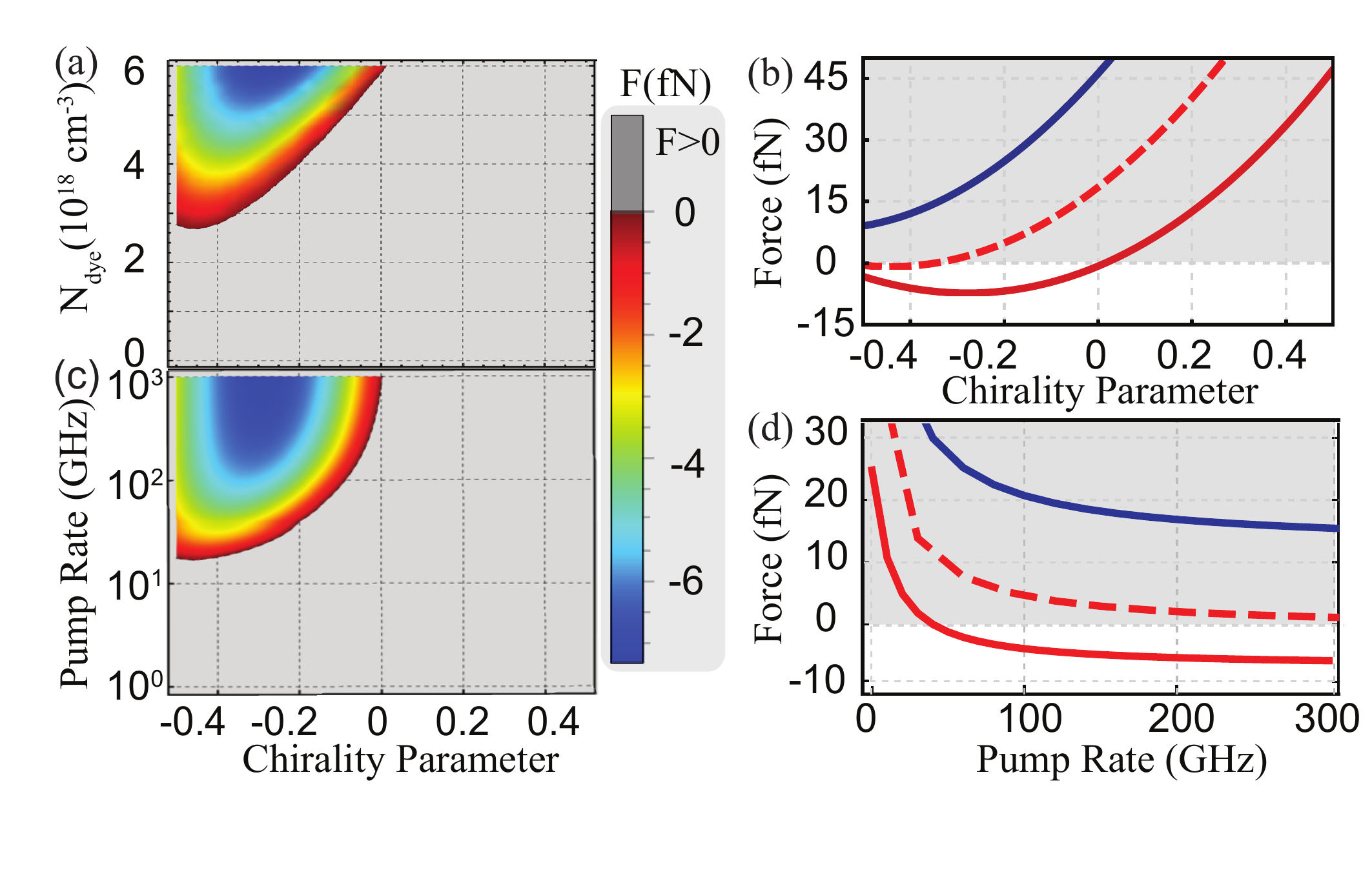}
\caption{(a) Net optical force as a function of chirality parameter and number density $N_{dye}$. (b) Optical force versus chirality parameter for $N_{dye}=0$ (blue), $N_{dye}=3\times10^{18}{\rm cm^{-3}}$ (dashed) and $N_{dye}=6\times10^{18}{\rm cm^{-3}}$ (red). (c) Force density plot as a function of chirality parameter and pump rate for fixed $N_{dye}=5\times10^{18}{\rm cm^{-3}}$. (d) Optical force as a function of pump rate for different chirality parameters :{ $\kappa = 0.2$ (blue), ${ \kappa = {0} }$ (dashed) and $\kappa= -0.2$ (red)}. In all calculation $b=200 {\rm \,nm}$, $t=80{\rm\, nm}$, $\sigma=+1$ and $\lambda=1064{\rm\, nm}.$ } 
 \label{F3}
\end{figure}

 We now demonstrate the tunability of the enantioselection by taking moderate values of dye-concentration \cite{Chen2005} and experimentally achievable pump power, where dye quenching and nonlinear effect can be ignored \cite{Pezzi2019,Polimeno2020}. In Fig. \ref{F3}(a), we compute the optical force acting on a dye-enriched core-shell particle as a function of the chirality parameter and $N_\text{dye}$ for fixed $\Gamma_\text{pump} =$\SI{ 100}{\giga\hertz}. Figure \ref{F3}(a) shows that optical pulling forces ($F<0$) gradually vanishes as the chirality parameter approaches zero, showing that pulling force is not possible for non-chiral gain-enhanced particles at this dye concentration. Also, the optical force is always pushing ($F>0$) for passive spheres ($N_{dye}$=0) regardless of the chirality parameter. 
 As one increases $N_{dye}$ the emission photons by the dye molecules in the core also increase and since the scattering force $F_s$ for right-handed chiral $\kappa<0$ particle is weak, there is a critical value $N_{dye} \approx 3.2\times10^{18}{ \rm cm^{-3}}$, above which right-handed chiral particles experiences pulling forces. On the other hand, for left-handed chiral $\kappa>0$ spheres, the optical force is always pushing ($F>0$) due to large $F_s$. This crossover between optical pulling and pushing forces for particles with $\kappa<0$ shells can also be noticed by inspecting horizontal linecuts of the density plots shown in Fig. \ref{F3}(a), as shown in Fig. \ref{F3}(b), where $F$ as a function of $\kappa$ is calculated for increasing dye concentrations $N_{dye}=\{0,3,6\}\times10^{18}\rm cm^{-3}$, for fixed $\Gamma = 100$ GHz. Interestingly, for $N_\text{dye}>6\times10^{18} \rm cm^{-3}$, chiral resolution is achieved for particle with very small shell chirality parameter, suggesting that that the proposed enantioselective mechanism could also be applied to naturally occurring materials, which typically have smaller values of $\kappa$. Although tuning the dye concentration might not be easily accomplished, Eq.~\ref{neff} shows that varying the pump rate amplification can rate at a fixed dye concentration is also feasible. This tuning knob is explored in Fig. \ref{F3}(c,d), the net optical force is calculated as a function of the chirality parameter and $\Gamma_{\text{pump}}$ for fixed $N_\text{dye} = 5 \times10^{18} {\rm cm^{-3}} $. The behavior is similar to varying dye concentration, a power threshold exists to achieve pulling forces, provided some degree of chirality is present.

\begin{figure}
\centering
\includegraphics[scale=.88]{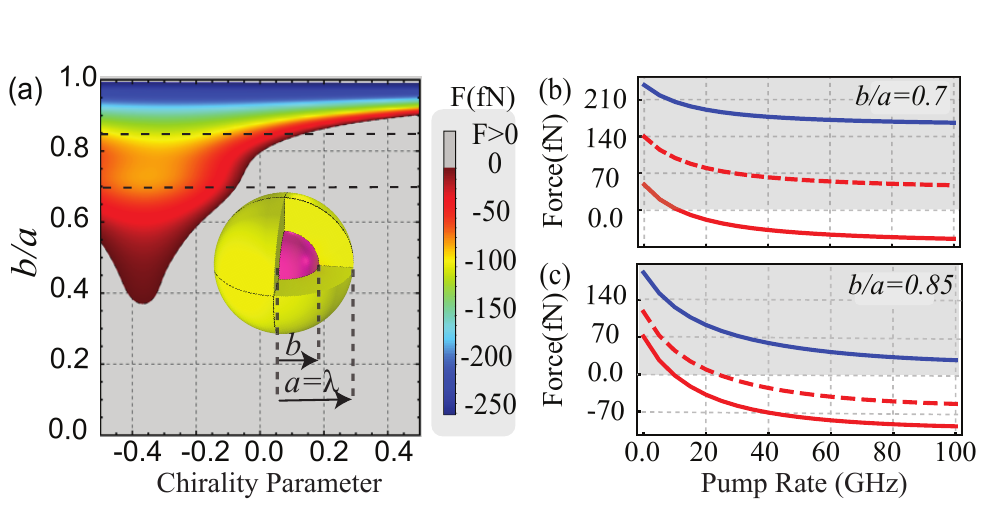}
\caption{
(a) Density plot representing the 
optical force as a function of the chirality parameter and the ratio between the core radius $b$ to out radius $a = b+t= \lambda$ for fixed $\Gamma_{\text{pump}}$ = \SI{ 100}{\giga\hertz} and $N_{dye}=5\times10^{18} {\rm\ cm^{-3}}$. Inset illustrates the particle under consideration. Optical force as a function of pump rate for fixed $b =0.7 a$ (b) and $b=0.85 a$ (c) for different chirality parameter: $\kappa=0.2$ (blue), $\kappa=0$ (dashed) and $\kappa=-0.2$ (red).  All the other parameters are the same as in figure (a). } 
 \label{F4}
\end{figure}

 It is worth mentioning that even for low incident intensity, $I_0=$\SI{1}{\mW\per\micro\meter^2}, the optical pulling force is 
 larger than the Brownian force. In the present study, the Brownian force $F_B \approx5\rm{\ fN}$ (see appendix \ref{app:1}). Although the pulling force can further be improved by increasing the pump rate or incident intensity $I_0$.
 To infer the critical pump rate needed to achieve to enantioselection, we show in Fig. \ref{F3}(d) vertical linecuts of Fig. \ref{F3}(c), revealing the optical force as a function of pumping rate for chirality parameters $\kappa=\{ -0.2, 0, +0.2\}$ for fixed ${ N_\text{dye} = 5 \times 10^{18}\,{\rm cm^{-3}}}$. While the optical force is always positive when $\Gamma_{\text{pump}}=0$, as the system is excited with an external optical pump, the sign of the net optical force is strongly dependent of the handedness of the chiral shell. In particular, the value of the chirality parameter of the shell defines the threshold pump rate at which the optical force changes sign, a consequence of the amplified photon emission, $F_a$ overcoming $F_s$, which is due to the scattered and radiative losses. This threshold pumping rate can also be estimated from Fig.\ref{F4}(c). 

To demonstrate the robustness of our findings against the geometrical parameters of the core-shell nanostructure, we calculate the optical force as a function of the chirality parameter and the ratio between the core $b$ to outer radii $a=b+t$ in Fig. \ref{F4}(a).
It is worth noting that for small active core $b/a\ll 1$ where the role of dye molecules is negligible so that $F_s\gg F_a$ and, therefore, the total optical force is positive. At $b=0.4 a$ the core/shell with $\kappa\approx-0.4$ undergoes a pushing to pulling force crossover due to the fact that in this situation the particle is expected to be subjected to weak scattering force $F_s$; hence a small emission of photons is sufficient to overcome the scattering force $F_a>F_s$. As a result  optical pulling force exists for $\kappa = -0.4$ while pushing forces occur for $\kappa=0.4$. This result shows that an enantioselective mechanism could be implemented for core-shell spheres with a broad range of geometrical parameters. On the other hand, for $t \gg b$,  when  active core is large, the particle is subjected to pulling force regardless of the chirality parameters.

Similarly to the previous analysis, the threshold pump power required to perform chiral resolution can be more clearly seen by inspecting the force dependence on the pump rate. 
The robustness of the chiral resolution to geometry fluctuations can be noticed by  considering two geometric configurations: $b = 0.7 a$ and $b = 0.85a$; the corresponding optical forces, as a function of the pump rate, are plotted in \ref{F4}(b) and \ref{F4}(c), respectively, for different values of the chirality parameter: { $\kappa = 0.2 $ (blue), $\kappa = 0.0 $ (dashed) and $\kappa =-0.2 $ (red)}.
 From the analysis of Fig. \ref{F4}(b) and \ref{F4}(c) one can conclude that right-handed chiral spheres undergo pulling force for pump rate larger than \SI{ 10}{\giga\hertz}, despite the fact these two systems have different gain levels the different sizes of the active gain core. 
In order to understand this result, one should notice that the scattering force is smaller in the configuration depicted in Fig. \ref{F4}(b), which has a thicker chiral shell. As a result for such a system, a weaker light amplification is sufficient to fulfill the condition $|F_\text{a}|>|F_\text{s}|$ to achieve a net optical pulling force. 
On the other hand, in comparison to the configuration shown in Fig. \ref{F4}(c) the shell is thinner and $|F_\text{s}|$ is expected to be larger, but at the same time $|F_\text{a}|$ becomes large due to the increased  volume of the gain media. This fact compensates the increase of $|F_\text{s}|$ for generating a net pulling force in the configuration of Fig. \ref{F4}(c). Consequently, the overall result is that such configuration exhibits approximately the same critical pump rate (\SI{ 25}{\giga\hertz}) to allow for pulling forces.
 
 \begin{figure}
     \centering
     \includegraphics[scale=0.5]{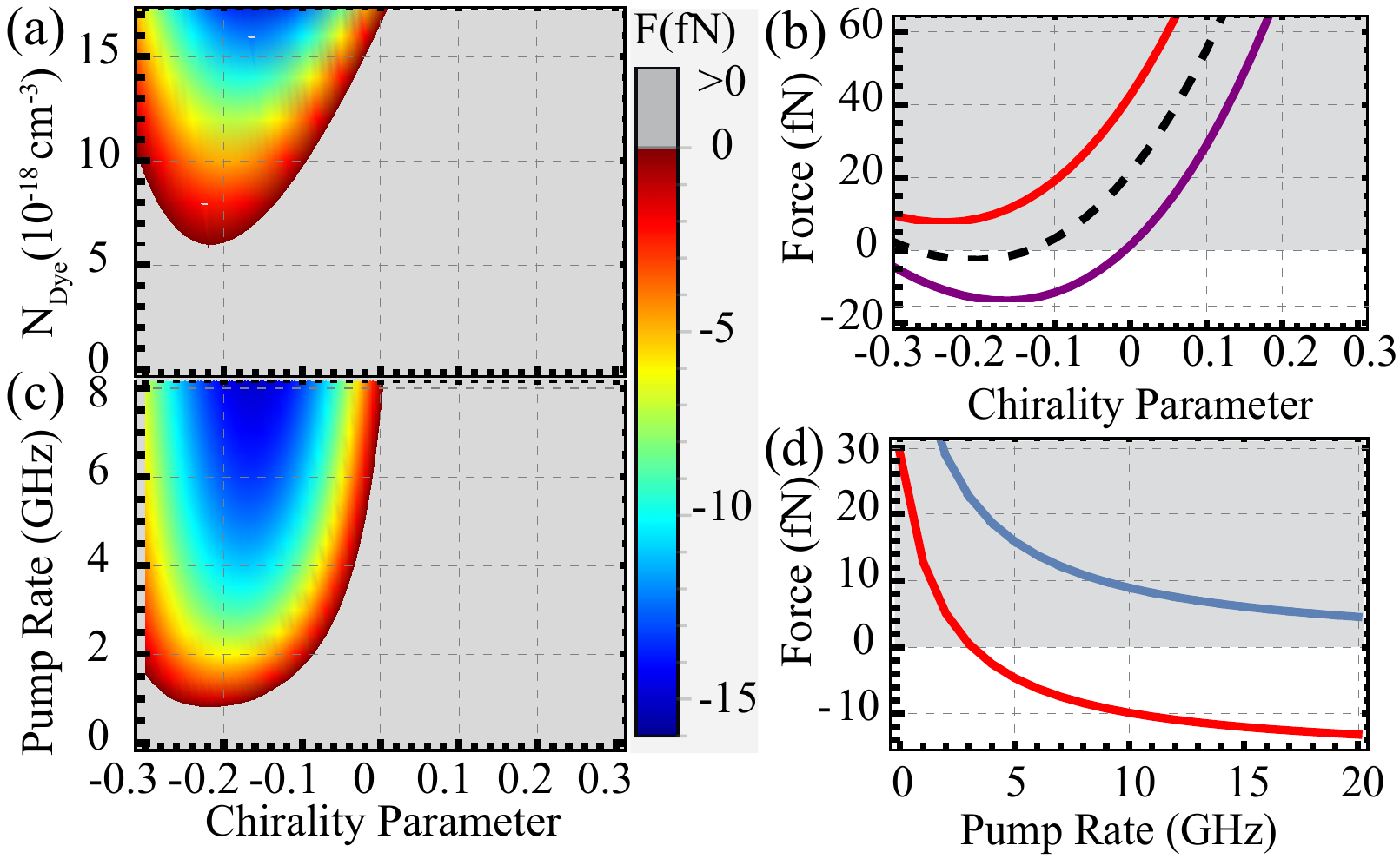}
     \caption{ (a) Optical force acting on a gain functionalized chiral sphere as a function of a chirality parameter $\kappa$ and dye concentration for fixed pump rate at 50 GHz. (b) Optical force versus chirality parameter for different dye concentration: $N_{dye}=0$  (red),  $N_{dye}=8\times10^{18} {\rm\ cm}^{-3}$  (dotted) and  $N_{dye}=16\times10^{18}{\rm\ cm}^{-3}$ (purple), where pumping rate is the same as (a). 
     (c) Optical force as a function of  chirality parameter  and  pumping rate for fixed dye concentration at $1.5\times10^{18} {\rm\ cm}^{-3}$.
     The optical force acting on a sphere with chirality parameter: $\kappa=-5\times 10^{-2}$ (red) and $\kappa=5\times 10^{-2}$ (darker) as a function of the pup rate, where dye concentration is fixed at $N=1.5\times 10^{18}{\rm\ cm}^{-3}$. This calculation is carried out by taking the refractive index of the sphere $n_s= \sqrt{\epsilon_g}+\sigma \kappa$, where $\epsilon_g$ is given previous section. }
     \label{Fsupp}
 \end{figure}
\subsection*{Optical pulling force on a homogeneous chiral sphere}
 In order to improve the generality of this scheme, in this section we discuss optical forces on a gain-functionalized homogeneous chiral sphere. This type of chiral sphere may constitute a variety of the synthesized chiral example such as DNA-assembled nanoparticles,  Cholesteric liquid crystals, carbon nanotubes, chiral nanocrystals, and chiral
quantum dots \cite{Urban2019,fan2010}. In addition to the spherical particles, this approach can also be applied to asymmetric or random shape geometrical particles  that show large circular dichroism as shown in \cite{fan2010,fan2012chiral,Cathcart2011}.    Without losing the generality, we  have added the dye molecules inside the chiral sphere to activate the role of gain media (see details in the previous section). As a continuation of our discussion reported in the main text, we calculate the optical force acting on the active chiral sphere as a function of the chirality parameter and dye concentration in Fig. \ref{Fsupp}(a). The results show that even for this configuration we can exert chirality-dependent optical forces on the sphere which provides an opportunity to perform  chiral selection of the homogeneous chiral sphere. For deeper insights, we consider a chiral sphere with an arbitrarily small value of $\kappa = \pm 0.05$ illuminated by a left-handed circularly polarized field to calculate the optical force as a function pump rate for fixed dye concentration.
 Figure  \ref{Fsupp}(b) shows that even for small chirality parameters right-handed chiral experiences pulling force $F\approx$\SI{10}{\femto\newton} for incident intensity $I_0=$\SI{1}{\mW\per\micro\meter^2}. This confirms that this is a sufficiently large force to overcome the Brownian force. 
 Thus, our scheme is robust against the selection of the chiral structures and  one can perform the chiral resolution of the racemic mixture by controlling the pumping rate of the dye molecules for small values of the chirality parameter. 

\subsection*{Pulling force on chiral dipoles}
\begin{figure}
\centering
\includegraphics[scale=.48]{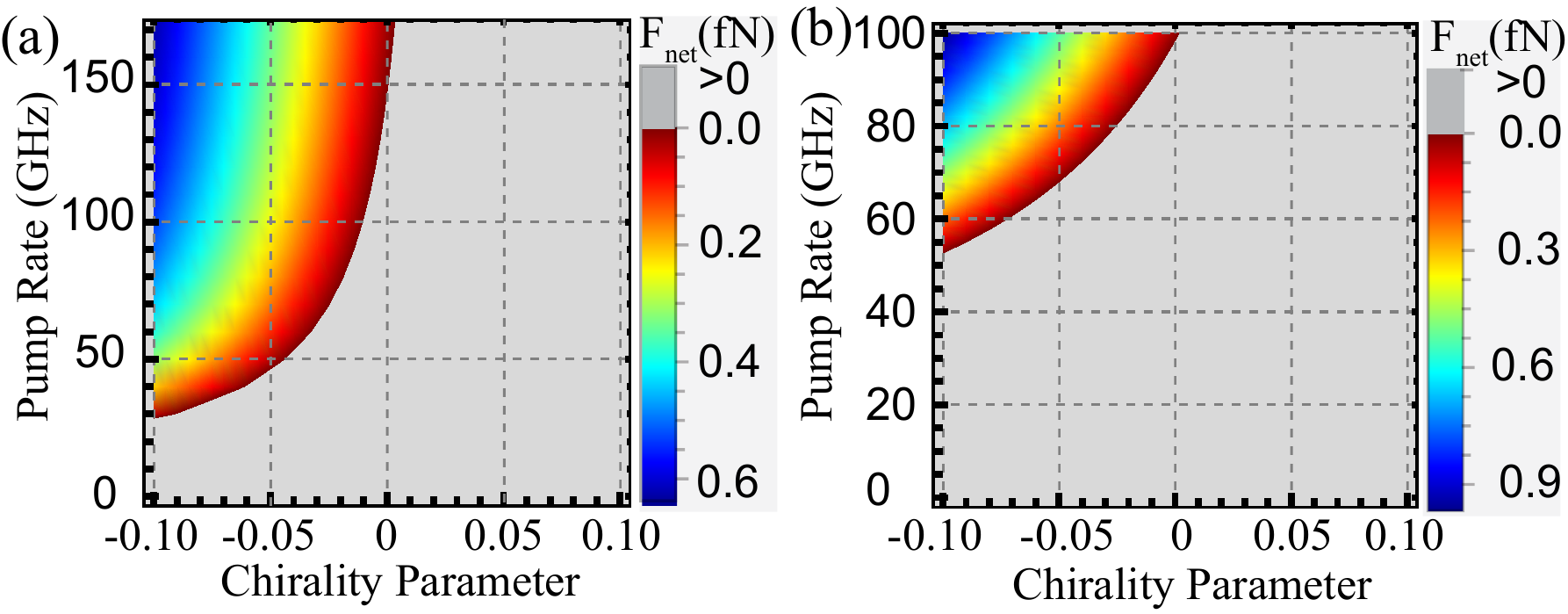}
\caption{
(a) Density plot representing the 
net optical force  $F_{net} (fN)$ as a function of chirality parameter and  $\Gamma_{\text{pump}}$  for fixed (a) $N_{dye}=6.4\times10^{18} {\rm cm^{-3}}$ and (b) $N_{dye}=0.3\times10^{18} {\rm cm^{-3}}$. The color region shows the stable optical pulling force, where we have subtracted the Brownian force from the optical force. } 
 \label{F5}
\end{figure}

In this study we have shown that the  gain-functionalized chiral spheres can be optically sorted using single-plane waves regardless of size and particle choices. To extend the applicability of our proposal to chiral dipoles, where only electric dipoles are prominent \cite{Ali2020Tailoring}, we consider a core-shell chiral sphere or radius $50 {\rm nm}$ (with active $20 {\rm nm}$ and chiral shell $30 {\rm nm}$) in Fig. \ref{F5}(a) and  a homogeneous chiral sphere of radius $50 {\rm nm}$ in Fig. \ref{F5}(b). Finally, we calculate the density plot of the  stable optical pulling force ($F_{net}=F-F_B$) acting on the spheres as a function of the chirality parameter and pump power for fixed dye concentration. 
Thus, our approach remains valid for small particles with weak chirality.

\section{Conclusion}
 In conclusion, we demonstrate an externally controllable and robust enantioselective method for single, isolated chiral particles doped with gain material.
 By considering realistic material and geometrical parameters,
 we show that this system allows one to perform enantioselection not only for lossless chiral media but also for plasmonic chiral spheres using single-plane waves. 
It is demonstrated that by externally tuning the pump rate, a crossover between pulling-to-pushing force occurs  for chiral shells with a given handedness, allowing for enantioselection of a racemic mixture of core-shell particles. We also demonstrate that this chiral resolution method is robust against varying the geometric parameters of the system and may be applied for arbitrarily small values of the chiral parameter, suggesting that one can also apply it to naturally occurring chiral materials.

\subsection*{ Funding} 
This work was supported by S\~ao Paulo Research Foundation (FAPESP) through grants 2020/03131-2, 2018/15580-6, 2018/15577-5, 2018/25339-4, Coordena{\c c}\~ao de Aperfei{\c c}oamento de Pessoal de N{\'i}vel Superior - Brasil (CAPES) (Finance Code 001), Conselho Nacional de Desenvolvimento Científico e Tecnológico through grants 425338/2018-5, 310224/2018-7, and Financiadora de Estudos e Projetos (Finep). F.A.P thanks financial support CNPq, CAPES, and FAPERJ. 
\subsection*{ Disclosures} The authors declare no conflicts of interest. Additional data and simulation files used in this article are available at \href{http://www.doi.org10.5281/zenodo.7841096}{10.5281/zenodo.7841096} upon publication.

\appendix
\section{Brownian Motion\label{app:1}}
A sphere immersed in the aqueous solution (water in our case) must be subjected to the Brownian force according to the fluctuation-dissipation theorem, which can be expressed as $F_B =\sqrt{12\pi \eta a k_\beta T}$ \cite{Okamoto1999}. Here, $a$ is the sphere radius, $\eta$ is fluid viscosity, $k_\beta$ is Boltzmann constant and T is room temperature. For the case of a sub-wavelength particle, say $a=200 nm$, immersed in the water of viscosity $\eta=7.9\times10^{-4}\rm{\ Pa.s}$ \cite{Li2016,Okamoto1999}, the Brownian force acting on the sphere is turned out as $F_B = 5{\rm\ fN}$ \cite{Li2016}. However, our results show that optical forces acting on the sphere are larger than the Brownian force even for weak incident intensity $I_0=$\SI{1}{\mW\per\micro\meter^2}~\cite{Brzobohaty2013}. Thus, the optical forces acting on the particle can easily overcome the thermal and Brownian fluctuations. In addition, optical forces can also be improved by increasing the power of the probing laser without changing the gain level as suggested by force expression (e.g.  $F\propto I_0$).
\bibliography{sample}

\begin{thebibliography}{41}%
\makeatletter
\providecommand \@ifxundefined [1]{%
 \@ifx{#1\undefined}
}%
\providecommand \@ifnum [1]{%
 \ifnum #1\expandafter \@firstoftwo
 \else \expandafter \@secondoftwo
 \fi
}%
\providecommand \@ifx [1]{%
 \ifx #1\expandafter \@firstoftwo
 \else \expandafter \@secondoftwo
 \fi
}%
\providecommand \natexlab [1]{#1}%
\providecommand \enquote  [1]{``#1''}%
\providecommand \bibnamefont  [1]{#1}%
\providecommand \bibfnamefont [1]{#1}%
\providecommand \citenamefont [1]{#1}%
\providecommand \href@noop [0]{\@secondoftwo}%
\providecommand \href [0]{\begingroup \@sanitize@url \@href}%
\providecommand \@href[1]{\@@startlink{#1}\@@href}%
\providecommand \@@href[1]{\endgroup#1\@@endlink}%
\providecommand \@sanitize@url [0]{\catcode `\\12\catcode `\$12\catcode
  `\&12\catcode `\#12\catcode `\^12\catcode `\_12\catcode `\%12\relax}%
\providecommand \@@startlink[1]{}%
\providecommand \@@endlink[0]{}%
\providecommand \url  [0]{\begingroup\@sanitize@url \@url }%
\providecommand \@url [1]{\endgroup\@href {#1}{\urlprefix }}%
\providecommand \urlprefix  [0]{URL }%
\providecommand \Eprint [0]{\href }%
\providecommand \doibase [0]{https://doi.org/}%
\providecommand \selectlanguage [0]{\@gobble}%
\providecommand \bibinfo  [0]{\@secondoftwo}%
\providecommand \bibfield  [0]{\@secondoftwo}%
\providecommand \translation [1]{[#1]}%
\providecommand \BibitemOpen [0]{}%
\providecommand \bibitemStop [0]{}%
\providecommand \bibitemNoStop [0]{.\EOS\space}%
\providecommand \EOS [0]{\spacefactor3000\relax}%
\providecommand \BibitemShut  [1]{\csname bibitem#1\endcsname}%
\let\auto@bib@innerbib\@empty
\bibitem [{\citenamefont {Wagni\`eree}(2007)}]{Wagniere2007}%
  \BibitemOpen
  \bibfield  {author} {\bibinfo {author} {\bibfnamefont {G.~H.}\ \bibnamefont
  {Wagni\`eree}},\ }\href@noop {} {\emph {\bibinfo {title} {On chirality and
  the universal asymmetry : reflections on image and mirror image}}}\ (\bibinfo
   {publisher} {VHCA [with] Wiley-VCH},\ \bibinfo {address} {Z\:urich},\
  \bibinfo {year} {2007})\BibitemShut {NoStop}%
\bibitem [{\citenamefont {Hentschel}\ \emph {et~al.}(2017)\citenamefont
  {Hentschel}, \citenamefont {Schäferling}, \citenamefont {Duan},
  \citenamefont {Giessen},\ and\ \citenamefont {Liu}}]{Hentschel2017}%
  \BibitemOpen
  \bibfield  {author} {\bibinfo {author} {\bibfnamefont {M.}~\bibnamefont
  {Hentschel}}, \bibinfo {author} {\bibfnamefont {M.}~\bibnamefont
  {Schäferling}}, \bibinfo {author} {\bibfnamefont {X.}~\bibnamefont {Duan}},
  \bibinfo {author} {\bibfnamefont {H.}~\bibnamefont {Giessen}},\ and\ \bibinfo
  {author} {\bibfnamefont {N.}~\bibnamefont {Liu}},\ }\bibfield  {title}
  {\bibinfo {title} {Chiral plasmonics},\ }\href
  {https://doi.org/10.1126/sciadv.1602735} {\bibfield  {journal} {\bibinfo
  {journal} {Science Advances}\ }\textbf {\bibinfo {volume} {3}},\ \bibinfo
  {pages} {e1602735} (\bibinfo {year} {2017})},\ \Eprint
  {https://arxiv.org/abs/https://www.science.org/doi/pdf/10.1126/sciadv.1602735}
  {https://www.science.org/doi/pdf/10.1126/sciadv.1602735} \BibitemShut
  {NoStop}%
\bibitem [{\citenamefont {Fan}\ and\ \citenamefont {Govorov}(2010)}]{fan2010}%
  \BibitemOpen
  \bibfield  {author} {\bibinfo {author} {\bibfnamefont {Z.}~\bibnamefont
  {Fan}}\ and\ \bibinfo {author} {\bibfnamefont {A.~O.}\ \bibnamefont
  {Govorov}},\ }\bibfield  {title} {\bibinfo {title} {Plasmonic circular
  dichroism of chiral metal nanoparticle assemblies},\ }\bibfield  {booktitle}
  {\emph {\bibinfo {booktitle} {Nano Letters}},\ }\href
  {https://doi.org/10.1021/nl101231b} {\bibfield  {journal} {\bibinfo
  {journal} {Nano Letters}\ }\textbf {\bibinfo {volume} {10}},\ \bibinfo
  {pages} {2580} (\bibinfo {year} {2010})}\BibitemShut {NoStop}%
\bibitem [{\citenamefont {Kuzyk}\ \emph {et~al.}(2012)\citenamefont {Kuzyk},
  \citenamefont {Schreiber}, \citenamefont {Fan}, \citenamefont {Pardatscher},
  \citenamefont {Roller}, \citenamefont {H{\"o}gele}, \citenamefont {Simmel},
  \citenamefont {Govorov},\ and\ \citenamefont {Liedl}}]{kuzyk2012}%
  \BibitemOpen
  \bibfield  {author} {\bibinfo {author} {\bibfnamefont {A.}~\bibnamefont
  {Kuzyk}}, \bibinfo {author} {\bibfnamefont {R.}~\bibnamefont {Schreiber}},
  \bibinfo {author} {\bibfnamefont {Z.}~\bibnamefont {Fan}}, \bibinfo {author}
  {\bibfnamefont {G.}~\bibnamefont {Pardatscher}}, \bibinfo {author}
  {\bibfnamefont {E.-M.}\ \bibnamefont {Roller}}, \bibinfo {author}
  {\bibfnamefont {A.}~\bibnamefont {H{\"o}gele}}, \bibinfo {author}
  {\bibfnamefont {F.~C.}\ \bibnamefont {Simmel}}, \bibinfo {author}
  {\bibfnamefont {A.~O.}\ \bibnamefont {Govorov}},\ and\ \bibinfo {author}
  {\bibfnamefont {T.}~\bibnamefont {Liedl}},\ }\bibfield  {title} {\bibinfo
  {title} {Dna-based self-assembly of chiral plasmonic nanostructures with
  tailored optical response},\ }\href {https://doi.org/10.1038/nature10889}
  {\bibfield  {journal} {\bibinfo  {journal} {Nature}\ }\textbf {\bibinfo
  {volume} {483}},\ \bibinfo {pages} {311} (\bibinfo {year}
  {2012})}\BibitemShut {NoStop}%
\bibitem [{\citenamefont {Zhang}\ \emph {et~al.}(2005)\citenamefont {Zhang},
  \citenamefont {Albelda}, \citenamefont {Liu},\ and\ \citenamefont
  {Canary}}]{zhang2005}%
  \BibitemOpen
  \bibfield  {author} {\bibinfo {author} {\bibfnamefont {J.}~\bibnamefont
  {Zhang}}, \bibinfo {author} {\bibfnamefont {M.~T.}\ \bibnamefont {Albelda}},
  \bibinfo {author} {\bibfnamefont {Y.}~\bibnamefont {Liu}},\ and\ \bibinfo
  {author} {\bibfnamefont {J.~W.}\ \bibnamefont {Canary}},\ }\bibfield  {title}
  {\bibinfo {title} {Chiral nanotechnology},\ }\href
  {https://doi.org/https://doi.org/10.1002/chir.20178} {\bibfield  {journal}
  {\bibinfo  {journal} {Chirality}\ }\textbf {\bibinfo {volume} {17}},\
  \bibinfo {pages} {404} (\bibinfo {year} {2005})}\BibitemShut {NoStop}%
\bibitem [{\citenamefont {W.~H.~Brooks}\ and\ \citenamefont
  {G.~Daniel}(2011)}]{Brooks2011}%
  \BibitemOpen
  \bibfield  {author} {\bibinfo {author} {\bibfnamefont {C.~W.~G.}\
  \bibnamefont {W.~H.~Brooks}}\ and\ \bibinfo {author} {\bibfnamefont
  {K.}~\bibnamefont {G.~Daniel}},\ }\bibfield  {title} {\bibinfo {title} {The
  significance of chirality in drug design and development},\ }\href
  {https://doi.org/10.2174/156802611795165098} {\bibfield  {journal} {\bibinfo
  {journal} {Current Topics in Medicinal Chemistry}\ }\textbf {\bibinfo
  {volume} {11}},\ \bibinfo {pages} {760} (\bibinfo {year} {2011})}\BibitemShut
  {NoStop}%
\bibitem [{\citenamefont {Urban}\ \emph {et~al.}(2019)\citenamefont {Urban},
  \citenamefont {Shen}, \citenamefont {Kong}, \citenamefont {Zhu},
  \citenamefont {Govorov}, \citenamefont {Wang}, \citenamefont {Hentschel},\
  and\ \citenamefont {Liu}}]{Urban2019}%
  \BibitemOpen
  \bibfield  {author} {\bibinfo {author} {\bibfnamefont {M.~J.}\ \bibnamefont
  {Urban}}, \bibinfo {author} {\bibfnamefont {C.}~\bibnamefont {Shen}},
  \bibinfo {author} {\bibfnamefont {X.-T.}\ \bibnamefont {Kong}}, \bibinfo
  {author} {\bibfnamefont {C.}~\bibnamefont {Zhu}}, \bibinfo {author}
  {\bibfnamefont {A.~O.}\ \bibnamefont {Govorov}}, \bibinfo {author}
  {\bibfnamefont {Q.}~\bibnamefont {Wang}}, \bibinfo {author} {\bibfnamefont
  {M.}~\bibnamefont {Hentschel}},\ and\ \bibinfo {author} {\bibfnamefont
  {N.}~\bibnamefont {Liu}},\ }\bibfield  {title} {\bibinfo {title} {Chiral
  plasmonic nanostructures enabled by bottom-up approaches},\ }\href
  {https://doi.org/10.1146/annurev-physchem-050317-021332} {\bibfield
  {journal} {\bibinfo  {journal} {Annual Review of Physical Chemistry}\
  }\textbf {\bibinfo {volume} {70}},\ \bibinfo {pages} {275} (\bibinfo {year}
  {2019})},\ \bibinfo {note} {pMID: 31112458},\ \Eprint
  {https://arxiv.org/abs/https://doi.org/10.1146/annurev-physchem-050317-021332}
  {https://doi.org/10.1146/annurev-physchem-050317-021332} \BibitemShut
  {NoStop}%
\bibitem [{\citenamefont {Zhao}\ \emph {et~al.}(2016)\citenamefont {Zhao},
  \citenamefont {Saleh},\ and\ \citenamefont {Dionne}}]{dionne2016}%
  \BibitemOpen
  \bibfield  {author} {\bibinfo {author} {\bibfnamefont {Y.}~\bibnamefont
  {Zhao}}, \bibinfo {author} {\bibfnamefont {A.~A.~E.}\ \bibnamefont {Saleh}},\
  and\ \bibinfo {author} {\bibfnamefont {J.~A.}\ \bibnamefont {Dionne}},\
  }\bibfield  {title} {\bibinfo {title} {Enantioselective optical trapping of
  chiral nanoparticles with plasmonic tweezers},\ }\bibfield  {booktitle}
  {\emph {\bibinfo {booktitle} {ACS Photonics}},\ }\href
  {https://doi.org/10.1021/acsphotonics.5b00574} {\bibfield  {journal}
  {\bibinfo  {journal} {ACS Photonics}\ }\textbf {\bibinfo {volume} {3}},\
  \bibinfo {pages} {304} (\bibinfo {year} {2016})}\BibitemShut {NoStop}%
\bibitem [{\citenamefont {Zhao}\ \emph {et~al.}(2017)\citenamefont {Zhao},
  \citenamefont {Saleh}, \citenamefont {van~de Haar}, \citenamefont {Baum},
  \citenamefont {Briggs}, \citenamefont {Lay}, \citenamefont {Reyes-Becerra},\
  and\ \citenamefont {Dionne}}]{Zhao2017}%
  \BibitemOpen
  \bibfield  {author} {\bibinfo {author} {\bibfnamefont {Y.}~\bibnamefont
  {Zhao}}, \bibinfo {author} {\bibfnamefont {A.~A.~E.}\ \bibnamefont {Saleh}},
  \bibinfo {author} {\bibfnamefont {M.~A.}\ \bibnamefont {van~de Haar}},
  \bibinfo {author} {\bibfnamefont {B.}~\bibnamefont {Baum}}, \bibinfo {author}
  {\bibfnamefont {J.~A.}\ \bibnamefont {Briggs}}, \bibinfo {author}
  {\bibfnamefont {A.}~\bibnamefont {Lay}}, \bibinfo {author} {\bibfnamefont
  {O.~A.}\ \bibnamefont {Reyes-Becerra}},\ and\ \bibinfo {author}
  {\bibfnamefont {J.~A.}\ \bibnamefont {Dionne}},\ }\bibfield  {title}
  {\bibinfo {title} {Nanoscopic control and quantification of enantioselective
  optical forces},\ }\href {https://doi.org/10.1038/nnano.2017.180} {\bibfield
  {journal} {\bibinfo  {journal} {Nature Nanotechnology}\ }\textbf {\bibinfo
  {volume} {12}},\ \bibinfo {pages} {1055} (\bibinfo {year}
  {2017})}\BibitemShut {NoStop}%
\bibitem [{\citenamefont {Tkachenko}\ and\ \citenamefont
  {Brasselet}(2014)}]{Tkachenko2014Optofluidic}%
  \BibitemOpen
  \bibfield  {author} {\bibinfo {author} {\bibfnamefont {G.}~\bibnamefont
  {Tkachenko}}\ and\ \bibinfo {author} {\bibfnamefont {E.}~\bibnamefont
  {Brasselet}},\ }\bibfield  {title} {\bibinfo {title} {Optofluidic sorting of
  material chirality by chiral light},\ }\href
  {https://doi.org/10.1038/ncomms4577} {\bibfield  {journal} {\bibinfo
  {journal} {Nature Communications}\ }\textbf {\bibinfo {volume} {5}},\
  \bibinfo {pages} {3577} (\bibinfo {year} {2014})}\BibitemShut {NoStop}%
\bibitem [{\citenamefont {Wang}\ and\ \citenamefont {Chan}(2014)}]{wang2014}%
  \BibitemOpen
  \bibfield  {author} {\bibinfo {author} {\bibfnamefont {S.~B.}\ \bibnamefont
  {Wang}}\ and\ \bibinfo {author} {\bibfnamefont {C.~T.}\ \bibnamefont
  {Chan}},\ }\bibfield  {title} {\bibinfo {title} {Lateral optical force on
  chiral particles near a surface},\ }\href
  {https://doi.org/10.1038/ncomms4307} {\bibfield  {journal} {\bibinfo
  {journal} {Nature Communications}\ }\textbf {\bibinfo {volume} {5}},\
  \bibinfo {pages} {3307} (\bibinfo {year} {2014})}\BibitemShut {NoStop}%
\bibitem [{\citenamefont {Chen}\ \emph {et~al.}(2016)\citenamefont {Chen},
  \citenamefont {Liang}, \citenamefont {Liu},\ and\ \citenamefont
  {Lin}}]{chen2016}%
  \BibitemOpen
  \bibfield  {author} {\bibinfo {author} {\bibfnamefont {H.}~\bibnamefont
  {Chen}}, \bibinfo {author} {\bibfnamefont {C.}~\bibnamefont {Liang}},
  \bibinfo {author} {\bibfnamefont {S.}~\bibnamefont {Liu}},\ and\ \bibinfo
  {author} {\bibfnamefont {Z.}~\bibnamefont {Lin}},\ }\bibfield  {title}
  {\bibinfo {title} {Chirality sorting using two-wave-interference--induced
  lateral optical force},\ }\href {https://doi.org/10.1103/PhysRevA.93.053833}
  {\bibfield  {journal} {\bibinfo  {journal} {Phys. Rev. A}\ }\textbf {\bibinfo
  {volume} {93}},\ \bibinfo {pages} {053833} (\bibinfo {year}
  {2016})}\BibitemShut {NoStop}%
\bibitem [{\citenamefont {Zhang}\ \emph {et~al.}(2017)\citenamefont {Zhang},
  \citenamefont {Mahdy}, \citenamefont {Liu}, \citenamefont {Teng},
  \citenamefont {Lim}, \citenamefont {Wang},\ and\ \citenamefont
  {Qiu}}]{zhang2017}%
  \BibitemOpen
  \bibfield  {author} {\bibinfo {author} {\bibfnamefont {T.}~\bibnamefont
  {Zhang}}, \bibinfo {author} {\bibfnamefont {M.~R.~C.}\ \bibnamefont {Mahdy}},
  \bibinfo {author} {\bibfnamefont {Y.}~\bibnamefont {Liu}}, \bibinfo {author}
  {\bibfnamefont {J.~H.}\ \bibnamefont {Teng}}, \bibinfo {author}
  {\bibfnamefont {C.~T.}\ \bibnamefont {Lim}}, \bibinfo {author} {\bibfnamefont
  {Z.}~\bibnamefont {Wang}},\ and\ \bibinfo {author} {\bibfnamefont {C.-W.}\
  \bibnamefont {Qiu}},\ }\bibfield  {title} {\bibinfo {title} {All-optical
  chirality-sensitive sorting via reversible lateral forces in interference
  fields},\ }\bibfield  {booktitle} {\emph {\bibinfo {booktitle} {ACS Nano}},\
  }\href {https://doi.org/10.1021/acsnano.7b01428} {\bibfield  {journal}
  {\bibinfo  {journal} {ACS Nano}\ }\textbf {\bibinfo {volume} {11}},\ \bibinfo
  {pages} {4292} (\bibinfo {year} {2017})}\BibitemShut {NoStop}%
\bibitem [{\citenamefont {Ali}\ \emph {et~al.}(2021{\natexlab{a}})\citenamefont
  {Ali}, \citenamefont {Dutra}, \citenamefont {Pinheiro},\ and\ \citenamefont
  {Neto}}]{Ali2021}%
  \BibitemOpen
  \bibfield  {author} {\bibinfo {author} {\bibfnamefont {R.}~\bibnamefont
  {Ali}}, \bibinfo {author} {\bibfnamefont {R.~S.}\ \bibnamefont {Dutra}},
  \bibinfo {author} {\bibfnamefont {F.~A.}\ \bibnamefont {Pinheiro}},\ and\
  \bibinfo {author} {\bibfnamefont {P.~A.~M.}\ \bibnamefont {Neto}},\
  }\bibfield  {title} {\bibinfo {title} {Enantioselection and chiral sorting of
  single microspheres using optical pulling forces},\ }\href
  {https://doi.org/10.1364/OL.419150} {\bibfield  {journal} {\bibinfo
  {journal} {Opt. Lett.}\ }\textbf {\bibinfo {volume} {46}},\ \bibinfo {pages}
  {1640} (\bibinfo {year} {2021}{\natexlab{a}})}\BibitemShut {NoStop}%
\bibitem [{\citenamefont {Zheng}\ \emph {et~al.}(2021)\citenamefont {Zheng},
  \citenamefont {Li}, \citenamefont {Chen},\ and\ \citenamefont
  {Lin}}]{Zheng2021Light}%
  \BibitemOpen
  \bibfield  {author} {\bibinfo {author} {\bibfnamefont {H.}~\bibnamefont
  {Zheng}}, \bibinfo {author} {\bibfnamefont {X.}~\bibnamefont {Li}}, \bibinfo
  {author} {\bibfnamefont {H.}~\bibnamefont {Chen}},\ and\ \bibinfo {author}
  {\bibfnamefont {Z.}~\bibnamefont {Lin}},\ }\bibfield  {title} {\bibinfo
  {title} {Selective transport of chiral particles by optical pulling forces},\
  }\href {https://doi.org/10.1364/OE.444627} {\bibfield  {journal} {\bibinfo
  {journal} {Opt. Express}\ }\textbf {\bibinfo {volume} {29}},\ \bibinfo
  {pages} {42684} (\bibinfo {year} {2021})}\BibitemShut {NoStop}%
\bibitem [{\citenamefont {Ali}\ \emph {et~al.}(2020{\natexlab{a}})\citenamefont
  {Ali}, \citenamefont {Pinheiro}, \citenamefont {Dutra}, \citenamefont
  {Rosa},\ and\ \citenamefont {Maia~Neto}}]{Ali2020nanoscale}%
  \BibitemOpen
  \bibfield  {author} {\bibinfo {author} {\bibfnamefont {R.}~\bibnamefont
  {Ali}}, \bibinfo {author} {\bibfnamefont {F.~A.}\ \bibnamefont {Pinheiro}},
  \bibinfo {author} {\bibfnamefont {R.~S.}\ \bibnamefont {Dutra}}, \bibinfo
  {author} {\bibfnamefont {F.~S.~S.}\ \bibnamefont {Rosa}},\ and\ \bibinfo
  {author} {\bibfnamefont {P.~A.}\ \bibnamefont {Maia~Neto}},\ }\bibfield
  {title} {\bibinfo {title} {Enantioselective manipulation of single chiral
  nanoparticles using optical tweezers},\ }\href
  {https://doi.org/10.1039/C9NR09736H} {\bibfield  {journal} {\bibinfo
  {journal} {Nanoscale}\ }\textbf {\bibinfo {volume} {12}},\ \bibinfo {pages}
  {5031} (\bibinfo {year} {2020}{\natexlab{a}})}\BibitemShut {NoStop}%
\bibitem [{\citenamefont {Li}\ \emph {et~al.}(2021)\citenamefont {Li},
  \citenamefont {Yan}, \citenamefont {Zhang}, \citenamefont {Chen},\ and\
  \citenamefont {Yao}}]{Manman2021}%
  \BibitemOpen
  \bibfield  {author} {\bibinfo {author} {\bibfnamefont {M.}~\bibnamefont
  {Li}}, \bibinfo {author} {\bibfnamefont {S.}~\bibnamefont {Yan}}, \bibinfo
  {author} {\bibfnamefont {Y.}~\bibnamefont {Zhang}}, \bibinfo {author}
  {\bibfnamefont {X.}~\bibnamefont {Chen}},\ and\ \bibinfo {author}
  {\bibfnamefont {B.}~\bibnamefont {Yao}},\ }\bibfield  {title} {\bibinfo
  {title} {Optical separation and discrimination of chiral particles by vector
  beams with orbital angular momentum},\ }\href
  {https://doi.org/10.1039/D1NA00530H} {\bibfield  {journal} {\bibinfo
  {journal} {Nanoscale Adv.}\ }\textbf {\bibinfo {volume} {3}},\ \bibinfo
  {pages} {6897} (\bibinfo {year} {2021})}\BibitemShut {NoStop}%
\bibitem [{\citenamefont {Hayat}\ \emph {et~al.}(2015)\citenamefont {Hayat},
  \citenamefont {Mueller},\ and\ \citenamefont {Capasso}}]{Hayat2015}%
  \BibitemOpen
  \bibfield  {author} {\bibinfo {author} {\bibfnamefont {A.}~\bibnamefont
  {Hayat}}, \bibinfo {author} {\bibfnamefont {J.~P.~B.}\ \bibnamefont
  {Mueller}},\ and\ \bibinfo {author} {\bibfnamefont {F.}~\bibnamefont
  {Capasso}},\ }\bibfield  {title} {\bibinfo {title} {Lateral chirality-sorting
  optical forces},\ }\href {https://doi.org/10.1073/pnas.1516704112} {\bibfield
   {journal} {\bibinfo  {journal} {Proceedings of the National Academy of
  Sciences}\ }\textbf {\bibinfo {volume} {112}},\ \bibinfo {pages} {13190}
  (\bibinfo {year} {2015})},\ \Eprint
  {https://arxiv.org/abs/https://www.pnas.org/content/112/43/13190.full.pdf}
  {https://www.pnas.org/content/112/43/13190.full.pdf} \BibitemShut {NoStop}%
\bibitem [{\citenamefont {Chen}\ \emph {et~al.}(2011)\citenamefont {Chen},
  \citenamefont {Ng}, \citenamefont {Lin},\ and\ \citenamefont
  {Chan}}]{Chen2011optical}%
  \BibitemOpen
  \bibfield  {author} {\bibinfo {author} {\bibfnamefont {J.}~\bibnamefont
  {Chen}}, \bibinfo {author} {\bibfnamefont {J.}~\bibnamefont {Ng}}, \bibinfo
  {author} {\bibfnamefont {Z.}~\bibnamefont {Lin}},\ and\ \bibinfo {author}
  {\bibfnamefont {C.~T.}\ \bibnamefont {Chan}},\ }\bibfield  {title} {\bibinfo
  {title} {Optical pulling force},\ }\href
  {https://doi.org/10.1038/nphoton.2011.153} {\bibfield  {journal} {\bibinfo
  {journal} {Nature Photonics}\ }\textbf {\bibinfo {volume} {5}},\ \bibinfo
  {pages} {531} (\bibinfo {year} {2011})}\BibitemShut {NoStop}%
\bibitem [{\citenamefont {Fan}\ and\ \citenamefont
  {Govorov}(2012)}]{fan2012chiral}%
  \BibitemOpen
  \bibfield  {author} {\bibinfo {author} {\bibfnamefont {Z.}~\bibnamefont
  {Fan}}\ and\ \bibinfo {author} {\bibfnamefont {A.~O.}\ \bibnamefont
  {Govorov}},\ }\bibfield  {title} {\bibinfo {title} {Chiral nanocrystals:
  plasmonic spectra and circular dichroism},\ }\href@noop {} {\bibfield
  {journal} {\bibinfo  {journal} {Nano letters}\ }\textbf {\bibinfo {volume}
  {12}},\ \bibinfo {pages} {3283} (\bibinfo {year} {2012})}\BibitemShut
  {NoStop}%
\bibitem [{\citenamefont {Lu}\ \emph {et~al.}(2018)\citenamefont {Lu},
  \citenamefont {Yang}, \citenamefont {Wang}, \citenamefont {Yam},
  \citenamefont {Yu},\ and\ \citenamefont {Chen}}]{Lu2018}%
  \BibitemOpen
  \bibfield  {author} {\bibinfo {author} {\bibfnamefont {J.~E.}\ \bibnamefont
  {Lu}}, \bibinfo {author} {\bibfnamefont {C.-H.}\ \bibnamefont {Yang}},
  \bibinfo {author} {\bibfnamefont {H.}~\bibnamefont {Wang}}, \bibinfo {author}
  {\bibfnamefont {C.}~\bibnamefont {Yam}}, \bibinfo {author} {\bibfnamefont
  {Z.-G.}\ \bibnamefont {Yu}},\ and\ \bibinfo {author} {\bibfnamefont
  {S.}~\bibnamefont {Chen}},\ }\bibfield  {title} {\bibinfo {title} {Plasmonic
  circular dichroism of vesicle-like nanostructures by the template-less
  self-assembly of achiral janus nanoparticles},\ }\href
  {https://doi.org/10.1039/C8NR05366A} {\bibfield  {journal} {\bibinfo
  {journal} {Nanoscale}\ }\textbf {\bibinfo {volume} {10}},\ \bibinfo {pages}
  {14586} (\bibinfo {year} {2018})}\BibitemShut {NoStop}%
\bibitem [{\citenamefont {Rao}\ \emph {et~al.}(2015)\citenamefont {Rao},
  \citenamefont {Wang}, \citenamefont {Li}, \citenamefont {Zhang},
  \citenamefont {Xu},\ and\ \citenamefont {Ding}}]{Rao2015}%
  \BibitemOpen
  \bibfield  {author} {\bibinfo {author} {\bibfnamefont {C.}~\bibnamefont
  {Rao}}, \bibinfo {author} {\bibfnamefont {Z.-G.}\ \bibnamefont {Wang}},
  \bibinfo {author} {\bibfnamefont {N.}~\bibnamefont {Li}}, \bibinfo {author}
  {\bibfnamefont {W.}~\bibnamefont {Zhang}}, \bibinfo {author} {\bibfnamefont
  {X.}~\bibnamefont {Xu}},\ and\ \bibinfo {author} {\bibfnamefont
  {B.}~\bibnamefont {Ding}},\ }\bibfield  {title} {\bibinfo {title} {Tunable
  optical activity of plasmonic dimers assembled by dna origami},\ }\href
  {https://doi.org/10.1039/C5NR01634G} {\bibfield  {journal} {\bibinfo
  {journal} {Nanoscale}\ }\textbf {\bibinfo {volume} {7}},\ \bibinfo {pages}
  {9147} (\bibinfo {year} {2015})}\BibitemShut {NoStop}%
\bibitem [{\citenamefont {Lan}\ and\ \citenamefont {Wang}(2016)}]{Lan2016}%
  \BibitemOpen
  \bibfield  {author} {\bibinfo {author} {\bibfnamefont {X.}~\bibnamefont
  {Lan}}\ and\ \bibinfo {author} {\bibfnamefont {Q.}~\bibnamefont {Wang}},\
  }\bibfield  {title} {\bibinfo {title} {Self-assembly of chiral plasmonic
  nanostructures},\ }\href
  {https://doi.org/https://doi.org/10.1002/adma.201600697} {\bibfield
  {journal} {\bibinfo  {journal} {Advanced Materials}\ }\textbf {\bibinfo
  {volume} {28}},\ \bibinfo {pages} {10499} (\bibinfo {year}
  {2016})}\BibitemShut {NoStop}%
\bibitem [{\citenamefont {Cipparrone}\ \emph {et~al.}(2011)\citenamefont
  {Cipparrone}, \citenamefont {Mazzulla}, \citenamefont {Pane}, \citenamefont
  {Hernandez},\ and\ \citenamefont {Bartolino}}]{Cipparrone2011}%
  \BibitemOpen
  \bibfield  {author} {\bibinfo {author} {\bibfnamefont {G.}~\bibnamefont
  {Cipparrone}}, \bibinfo {author} {\bibfnamefont {A.}~\bibnamefont
  {Mazzulla}}, \bibinfo {author} {\bibfnamefont {A.}~\bibnamefont {Pane}},
  \bibinfo {author} {\bibfnamefont {R.~J.}\ \bibnamefont {Hernandez}},\ and\
  \bibinfo {author} {\bibfnamefont {R.}~\bibnamefont {Bartolino}},\ }\bibfield
  {title} {\bibinfo {title} {Chiral self-assembled solid microspheres: A novel
  multifunctional microphotonic device},\ }\href
  {https://doi.org/https://doi.org/10.1002/adma.201102828} {\bibfield
  {journal} {\bibinfo  {journal} {Advanced Materials}\ }\textbf {\bibinfo
  {volume} {23}},\ \bibinfo {pages} {5773} (\bibinfo {year}
  {2011})}\BibitemShut {NoStop}%
\bibitem [{Boh(1998{\natexlab{a}})}]{Bohren}%
  \BibitemOpen
  \bibinfo {title} {Absorption and scattering of light by small particles},\
  in\ \href {https://doi.org/https://doi.org/10.1002/9783527618156.ch4} {\emph
  {\bibinfo {booktitle} {Absorption and Scattering of Light by Small
  Particles}}}\ (\bibinfo  {publisher} {John Wiley and Sons, Ltd},\ \bibinfo
  {year} {1998})\ Chap.~\bibinfo {chapter} {4}, pp.\ \bibinfo {pages}
  {83--122}\BibitemShut {NoStop}%
\bibitem [{\citenamefont {Mizrahi}\ and\ \citenamefont
  {Fainman}(2010)}]{Mizrahi2010}%
  \BibitemOpen
  \bibfield  {author} {\bibinfo {author} {\bibfnamefont {A.}~\bibnamefont
  {Mizrahi}}\ and\ \bibinfo {author} {\bibfnamefont {Y.}~\bibnamefont
  {Fainman}},\ }\bibfield  {title} {\bibinfo {title} {Negative radiation
  pressure on gain medium structures},\ }\href
  {https://doi.org/10.1364/OL.35.003405} {\bibfield  {journal} {\bibinfo
  {journal} {Opt. Lett.}\ }\textbf {\bibinfo {volume} {35}},\ \bibinfo {pages}
  {3405} (\bibinfo {year} {2010})}\BibitemShut {NoStop}%
\bibitem [{\citenamefont {Kudo}\ and\ \citenamefont
  {Ishihara}(2012)}]{Kudo2012}%
  \BibitemOpen
  \bibfield  {author} {\bibinfo {author} {\bibfnamefont {T.}~\bibnamefont
  {Kudo}}\ and\ \bibinfo {author} {\bibfnamefont {H.}~\bibnamefont
  {Ishihara}},\ }\bibfield  {title} {\bibinfo {title} {Proposed nonlinear
  resonance laser technique for manipulating nanoparticles},\ }\href
  {https://doi.org/10.1103/PhysRevLett.109.087402} {\bibfield  {journal}
  {\bibinfo  {journal} {Phys. Rev. Lett.}\ }\textbf {\bibinfo {volume} {109}},\
  \bibinfo {pages} {087402} (\bibinfo {year} {2012})}\BibitemShut {NoStop}%
\bibitem [{Boh(1998{\natexlab{b}})}]{Bohren8}%
  \BibitemOpen
  \bibinfo {title} {Absorption and scattering of light by small particles},\
  in\ \href {https://doi.org/https://doi.org/10.1002/9783527618156.ch8} {\emph
  {\bibinfo {booktitle} {Absorption and Scattering of Light by Small
  Particles}}}\ (\bibinfo  {publisher} {John Wiley and Sons, Ltd},\ \bibinfo
  {year} {1998})\ Chap.~\bibinfo {chapter} {8}, pp.\ \bibinfo {pages}
  {182--223}\BibitemShut {NoStop}%
\bibitem [{\citenamefont {Pezzi}\ \emph {et~al.}(2019)\citenamefont {Pezzi},
  \citenamefont {Iatì}, \citenamefont {Saija}, \citenamefont {De~Luca},\ and\
  \citenamefont {Maragò}}]{Pezzi2019}%
  \BibitemOpen
  \bibfield  {author} {\bibinfo {author} {\bibfnamefont {L.}~\bibnamefont
  {Pezzi}}, \bibinfo {author} {\bibfnamefont {M.~A.}\ \bibnamefont {Iatì}},
  \bibinfo {author} {\bibfnamefont {R.}~\bibnamefont {Saija}}, \bibinfo
  {author} {\bibfnamefont {A.}~\bibnamefont {De~Luca}},\ and\ \bibinfo {author}
  {\bibfnamefont {O.~M.}\ \bibnamefont {Maragò}},\ }\bibfield  {title}
  {\bibinfo {title} {Resonant coupling and gain singularities in
  metal/dielectric multishells: Quasi-static versus t-matrix calculations},\
  }\href {https://doi.org/10.1021/acs.jpcc.9b07489} {\bibfield  {journal}
  {\bibinfo  {journal} {The Journal of Physical Chemistry C}\ }\textbf
  {\bibinfo {volume} {123}},\ \bibinfo {pages} {29291} (\bibinfo {year}
  {2019})},\ \Eprint
  {https://arxiv.org/abs/https://doi.org/10.1021/acs.jpcc.9b07489}
  {https://doi.org/10.1021/acs.jpcc.9b07489} \BibitemShut {NoStop}%
\bibitem [{\citenamefont {Polimeno}\ \emph {et~al.}(2020)\citenamefont
  {Polimeno}, \citenamefont {Patti}, \citenamefont {Infusino}, \citenamefont
  {Sánchez}, \citenamefont {Iatì}, \citenamefont {Saija}, \citenamefont
  {Volpe}, \citenamefont {Maragò},\ and\ \citenamefont
  {Veltri}}]{Polimeno2020}%
  \BibitemOpen
  \bibfield  {author} {\bibinfo {author} {\bibfnamefont {P.}~\bibnamefont
  {Polimeno}}, \bibinfo {author} {\bibfnamefont {F.}~\bibnamefont {Patti}},
  \bibinfo {author} {\bibfnamefont {M.}~\bibnamefont {Infusino}}, \bibinfo
  {author} {\bibfnamefont {J.}~\bibnamefont {Sánchez}}, \bibinfo {author}
  {\bibfnamefont {M.~A.}\ \bibnamefont {Iatì}}, \bibinfo {author}
  {\bibfnamefont {R.}~\bibnamefont {Saija}}, \bibinfo {author} {\bibfnamefont
  {G.}~\bibnamefont {Volpe}}, \bibinfo {author} {\bibfnamefont {O.~M.}\
  \bibnamefont {Maragò}},\ and\ \bibinfo {author} {\bibfnamefont
  {A.}~\bibnamefont {Veltri}},\ }\bibfield  {title} {\bibinfo {title}
  {Gain-assisted optomechanical position locking of metal/dielectric nanoshells
  in optical potentials},\ }\href
  {https://doi.org/10.1021/acsphotonics.0c00213} {\bibfield  {journal}
  {\bibinfo  {journal} {ACS Photonics}\ }\textbf {\bibinfo {volume} {7}},\
  \bibinfo {pages} {1262} (\bibinfo {year} {2020})},\ \Eprint
  {https://arxiv.org/abs/https://doi.org/10.1021/acsphotonics.0c00213}
  {https://doi.org/10.1021/acsphotonics.0c00213} \BibitemShut {NoStop}%
\bibitem [{\citenamefont {Campione}\ \emph {et~al.}(2011)\citenamefont
  {Campione}, \citenamefont {Albani},\ and\ \citenamefont
  {Capolino}}]{Campione2011}%
  \BibitemOpen
  \bibfield  {author} {\bibinfo {author} {\bibfnamefont {S.}~\bibnamefont
  {Campione}}, \bibinfo {author} {\bibfnamefont {M.}~\bibnamefont {Albani}},\
  and\ \bibinfo {author} {\bibfnamefont {F.}~\bibnamefont {Capolino}},\
  }\bibfield  {title} {\bibinfo {title} {Complex modes and near-zero
  permittivity in 3d arrays of plasmonic nanoshells: loss compensation using
  gain [invited]},\ }\href {https://doi.org/10.1364/OME.1.001077} {\bibfield
  {journal} {\bibinfo  {journal} {Opt. Mater. Express}\ }\textbf {\bibinfo
  {volume} {1}},\ \bibinfo {pages} {1077} (\bibinfo {year} {2011})}\BibitemShut
  {NoStop}%
\bibitem [{\citenamefont {Doan}\ \emph {et~al.}(2017)\citenamefont {Doan},
  \citenamefont {Castillo}, \citenamefont {Bejjani}, \citenamefont {Nurekeyev},
  \citenamefont {Dzyuba}, \citenamefont {Gryczynski}, \citenamefont
  {Gryczynski},\ and\ \citenamefont {Raut}}]{Doan2017}%
  \BibitemOpen
  \bibfield  {author} {\bibinfo {author} {\bibfnamefont {H.}~\bibnamefont
  {Doan}}, \bibinfo {author} {\bibfnamefont {M.}~\bibnamefont {Castillo}},
  \bibinfo {author} {\bibfnamefont {M.}~\bibnamefont {Bejjani}}, \bibinfo
  {author} {\bibfnamefont {Z.}~\bibnamefont {Nurekeyev}}, \bibinfo {author}
  {\bibfnamefont {S.~V.}\ \bibnamefont {Dzyuba}}, \bibinfo {author}
  {\bibfnamefont {I.}~\bibnamefont {Gryczynski}}, \bibinfo {author}
  {\bibfnamefont {Z.}~\bibnamefont {Gryczynski}},\ and\ \bibinfo {author}
  {\bibfnamefont {S.}~\bibnamefont {Raut}},\ }\bibfield  {title} {\bibinfo
  {title} {Solvatochromic dye lds 798 as microviscosity and ph probe},\ }\href
  {https://doi.org/10.1039/C7CP05874H} {\bibfield  {journal} {\bibinfo
  {journal} {Phys. Chem. Chem. Phys.}\ }\textbf {\bibinfo {volume} {19}},\
  \bibinfo {pages} {29934} (\bibinfo {year} {2017})}\BibitemShut {NoStop}%
\bibitem [{\citenamefont {Ali}(2022)}]{Ali2022acs}%
  \BibitemOpen
  \bibfield  {author} {\bibinfo {author} {\bibfnamefont {R.}~\bibnamefont
  {Ali}},\ }\bibfield  {title} {\bibinfo {title} {Tunable anomalous scattering
  and negative asymmetry parameter in a gain-functionalized low refractive
  index sphere},\ }\bibfield  {booktitle} {\emph {\bibinfo {booktitle} {ACS
  Omega}},\ }\href {https://doi.org/10.1021/acsomega.1c05662} {\bibfield
  {journal} {\bibinfo  {journal} {ACS Omega}\ }\textbf {\bibinfo {volume}
  {7}},\ \bibinfo {pages} {2170} (\bibinfo {year} {2022})}\BibitemShut
  {NoStop}%
\bibitem [{\citenamefont {Shi}\ \emph {et~al.}(2022)\citenamefont {Shi},
  \citenamefont {Zhou}, \citenamefont {Liu}, \citenamefont {Nieto-Vesperinas},
  \citenamefont {Zhu}, \citenamefont {Hassanfiroozi}, \citenamefont {Liu},
  \citenamefont {Zhang}, \citenamefont {Tsai}, \citenamefont {Li},
  \citenamefont {Ding}, \citenamefont {Zhu}, \citenamefont {Yu}, \citenamefont
  {Mazzulla}, \citenamefont {Cipparrone}, \citenamefont {Wu}, \citenamefont
  {Chan},\ and\ \citenamefont {Qiu}}]{Shi2022}%
  \BibitemOpen
  \bibfield  {author} {\bibinfo {author} {\bibfnamefont {Y.}~\bibnamefont
  {Shi}}, \bibinfo {author} {\bibfnamefont {L.-M.}\ \bibnamefont {Zhou}},
  \bibinfo {author} {\bibfnamefont {A.~Q.}\ \bibnamefont {Liu}}, \bibinfo
  {author} {\bibfnamefont {M.}~\bibnamefont {Nieto-Vesperinas}}, \bibinfo
  {author} {\bibfnamefont {T.}~\bibnamefont {Zhu}}, \bibinfo {author}
  {\bibfnamefont {A.}~\bibnamefont {Hassanfiroozi}}, \bibinfo {author}
  {\bibfnamefont {J.}~\bibnamefont {Liu}}, \bibinfo {author} {\bibfnamefont
  {H.}~\bibnamefont {Zhang}}, \bibinfo {author} {\bibfnamefont {D.~P.}\
  \bibnamefont {Tsai}}, \bibinfo {author} {\bibfnamefont {H.}~\bibnamefont
  {Li}}, \bibinfo {author} {\bibfnamefont {W.}~\bibnamefont {Ding}}, \bibinfo
  {author} {\bibfnamefont {W.}~\bibnamefont {Zhu}}, \bibinfo {author}
  {\bibfnamefont {Y.~F.}\ \bibnamefont {Yu}}, \bibinfo {author} {\bibfnamefont
  {A.}~\bibnamefont {Mazzulla}}, \bibinfo {author} {\bibfnamefont
  {G.}~\bibnamefont {Cipparrone}}, \bibinfo {author} {\bibfnamefont {P.~C.}\
  \bibnamefont {Wu}}, \bibinfo {author} {\bibfnamefont {C.~T.}\ \bibnamefont
  {Chan}},\ and\ \bibinfo {author} {\bibfnamefont {C.-W.}\ \bibnamefont
  {Qiu}},\ }\bibfield  {title} {\bibinfo {title} {Superhybrid mode-enhanced
  optical torques on mie-resonant particles},\ }\bibfield  {booktitle} {\emph
  {\bibinfo {booktitle} {Nano Letters}},\ }\href
  {https://doi.org/10.1021/acs.nanolett.2c00050} {\bibfield  {journal}
  {\bibinfo  {journal} {Nano Letters}\ }\textbf {\bibinfo {volume} {22}},\
  \bibinfo {pages} {1769} (\bibinfo {year} {2022})}\BibitemShut {NoStop}%
\bibitem [{\citenamefont {Ali}\ \emph {et~al.}(2021{\natexlab{b}})\citenamefont
  {Ali}, \citenamefont {de~Sousa~Dutra}, \citenamefont {Pinheiro},\ and\
  \citenamefont {Neto}}]{Ali2021jopt}%
  \BibitemOpen
  \bibfield  {author} {\bibinfo {author} {\bibfnamefont {R.}~\bibnamefont
  {Ali}}, \bibinfo {author} {\bibfnamefont {R.}~\bibnamefont {de~Sousa~Dutra}},
  \bibinfo {author} {\bibfnamefont {F.~A.}\ \bibnamefont {Pinheiro}},\ and\
  \bibinfo {author} {\bibfnamefont {P.~A.~M.}\ \bibnamefont {Neto}},\
  }\bibfield  {title} {\bibinfo {title} {Gain-assisted optical tweezing of
  plasmonic and large refractive index microspheres},\ }\href
  {https://doi.org/10.1088/2040-8986/ac228f} {\bibfield  {journal} {\bibinfo
  {journal} {Journal of Optics}\ }\textbf {\bibinfo {volume} {23}},\ \bibinfo
  {pages} {115004} (\bibinfo {year} {2021}{\natexlab{b}})}\BibitemShut
  {NoStop}%
\bibitem [{\citenamefont {Chen}\ \emph {et~al.}(2005)\citenamefont {Chen},
  \citenamefont {Wang}, \citenamefont {Ye}, \citenamefont {Ni}, \citenamefont
  {Chan}, \citenamefont {Yang},\ and\ \citenamefont {Lo}}]{Chen2005}%
  \BibitemOpen
  \bibfield  {author} {\bibinfo {author} {\bibfnamefont {F.}~\bibnamefont
  {Chen}}, \bibinfo {author} {\bibfnamefont {J.}~\bibnamefont {Wang}}, \bibinfo
  {author} {\bibfnamefont {C.}~\bibnamefont {Ye}}, \bibinfo {author}
  {\bibfnamefont {W.}~\bibnamefont {Ni}}, \bibinfo {author} {\bibfnamefont
  {J.}~\bibnamefont {Chan}}, \bibinfo {author} {\bibfnamefont {Y.}~\bibnamefont
  {Yang}},\ and\ \bibinfo {author} {\bibfnamefont {D.}~\bibnamefont {Lo}},\
  }\bibfield  {title} {\bibinfo {title} {Near infrared distributed feedback
  lasers based on lds dye-doped zirconia-organically modified silicate channel
  waveguides},\ }\href {https://doi.org/10.1364/OPEX.13.001643} {\bibfield
  {journal} {\bibinfo  {journal} {Opt. Express}\ }\textbf {\bibinfo {volume}
  {13}},\ \bibinfo {pages} {1643} (\bibinfo {year} {2005})}\BibitemShut
  {NoStop}%
\bibitem [{\citenamefont {Cathcart}\ and\ \citenamefont
  {Kitaev}(2011)}]{Cathcart2011}%
  \BibitemOpen
  \bibfield  {author} {\bibinfo {author} {\bibfnamefont {N.}~\bibnamefont
  {Cathcart}}\ and\ \bibinfo {author} {\bibfnamefont {V.}~\bibnamefont
  {Kitaev}},\ }\bibfield  {title} {\bibinfo {title} {Monodisperse hexagonal
  silver nanoprisms: Synthesis via thiolate-protected cluster precursors and
  chiral, ligand-imprinted self-assembly},\ }\href
  {https://doi.org/10.1021/nn2023478} {\bibfield  {journal} {\bibinfo
  {journal} {ACS Nano}\ }\textbf {\bibinfo {volume} {5}},\ \bibinfo {pages}
  {7411} (\bibinfo {year} {2011})}\BibitemShut {NoStop}%
\bibitem [{\citenamefont {Ali}\ \emph {et~al.}(2020{\natexlab{b}})\citenamefont
  {Ali}, \citenamefont {Pinheiro}, \citenamefont {Dutra},\ and\ \citenamefont
  {Neto}}]{Ali2020Tailoring}%
  \BibitemOpen
  \bibfield  {author} {\bibinfo {author} {\bibfnamefont {R.}~\bibnamefont
  {Ali}}, \bibinfo {author} {\bibfnamefont {F.~A.}\ \bibnamefont {Pinheiro}},
  \bibinfo {author} {\bibfnamefont {R.~S.}\ \bibnamefont {Dutra}},\ and\
  \bibinfo {author} {\bibfnamefont {P.~A.~M.}\ \bibnamefont {Neto}},\
  }\bibfield  {title} {\bibinfo {title} {Tailoring optical pulling forces with
  composite microspheres},\ }\href
  {https://doi.org/10.1103/PhysRevA.102.023514} {\bibfield  {journal} {\bibinfo
   {journal} {Phys. Rev. A}\ }\textbf {\bibinfo {volume} {102}},\ \bibinfo
  {pages} {023514} (\bibinfo {year} {2020}{\natexlab{b}})}\BibitemShut
  {NoStop}%
\bibitem [{\citenamefont {Okamoto}\ and\ \citenamefont
  {Kawata}(1999)}]{Okamoto1999}%
  \BibitemOpen
  \bibfield  {author} {\bibinfo {author} {\bibfnamefont {K.}~\bibnamefont
  {Okamoto}}\ and\ \bibinfo {author} {\bibfnamefont {S.}~\bibnamefont
  {Kawata}},\ }\bibfield  {title} {\bibinfo {title} {Radiation force exerted on
  subwavelength particles near a nanoaperture},\ }\href
  {https://doi.org/10.1103/PhysRevLett.83.4534} {\bibfield  {journal} {\bibinfo
   {journal} {Phys. Rev. Lett.}\ }\textbf {\bibinfo {volume} {83}},\ \bibinfo
  {pages} {4534} (\bibinfo {year} {1999})}\BibitemShut {NoStop}%
\bibitem [{\citenamefont {Li}\ \emph {et~al.}(2016)\citenamefont {Li},
  \citenamefont {Yan}, \citenamefont {Yao}, \citenamefont {Liang},
  \citenamefont {Han},\ and\ \citenamefont {Zhang}}]{Li2016}%
  \BibitemOpen
  \bibfield  {author} {\bibinfo {author} {\bibfnamefont {M.}~\bibnamefont
  {Li}}, \bibinfo {author} {\bibfnamefont {S.}~\bibnamefont {Yan}}, \bibinfo
  {author} {\bibfnamefont {B.}~\bibnamefont {Yao}}, \bibinfo {author}
  {\bibfnamefont {Y.}~\bibnamefont {Liang}}, \bibinfo {author} {\bibfnamefont
  {G.}~\bibnamefont {Han}},\ and\ \bibinfo {author} {\bibfnamefont
  {P.}~\bibnamefont {Zhang}},\ }\bibfield  {title} {\bibinfo {title} {Optical
  trapping force and torque on spheroidal rayleigh particles with arbitrary
  spatial orientations},\ }\href {https://doi.org/10.1364/JOSAA.33.001341}
  {\bibfield  {journal} {\bibinfo  {journal} {J. Opt. Soc. Am. A}\ }\textbf
  {\bibinfo {volume} {33}},\ \bibinfo {pages} {1341} (\bibinfo {year}
  {2016})}\BibitemShut {NoStop}%
\bibitem [{\citenamefont {Brzobohat{\'y}}\ \emph {et~al.}(2013)\citenamefont
  {Brzobohat{\'y}}, \citenamefont {Kar{\'a}sek}, \citenamefont {{\v S}iler},
  \citenamefont {Chv{\'a}tal}, \citenamefont {{\v C}i{\v z}m{\'a}r},\ and\
  \citenamefont {Zem{\'a}nek}}]{Brzobohaty2013}%
  \BibitemOpen
  \bibfield  {author} {\bibinfo {author} {\bibfnamefont {O.}~\bibnamefont
  {Brzobohat{\'y}}}, \bibinfo {author} {\bibfnamefont {V.}~\bibnamefont
  {Kar{\'a}sek}}, \bibinfo {author} {\bibfnamefont {M.}~\bibnamefont {{\v
  S}iler}}, \bibinfo {author} {\bibfnamefont {L.}~\bibnamefont {Chv{\'a}tal}},
  \bibinfo {author} {\bibfnamefont {T.}~\bibnamefont {{\v C}i{\v z}m{\'a}r}},\
  and\ \bibinfo {author} {\bibfnamefont {P.}~\bibnamefont {Zem{\'a}nek}},\
  }\bibfield  {title} {\bibinfo {title} {Experimental demonstration of optical
  transport, sorting and self-arrangement using a `tractor beam'},\ }\href
  {https://doi.org/10.1038/nphoton.2012.332} {\bibfield  {journal} {\bibinfo
  {journal} {Nature Photonics}\ }\textbf {\bibinfo {volume} {7}},\ \bibinfo
  {pages} {123} (\bibinfo {year} {2013})}\BibitemShut {NoStop}%
\end{thebibliography}%
\end{document}